\documentclass{aa}
\usepackage{graphicx}
\usepackage{txfonts}
\usepackage{url}
\newcommand{\xmm}{{\em XMM-Newton}}

\newcommand{\cxo}{{\em Chandra}}

\newcommand{{\myr}}{\mbox{[$M_\odot\,{\rm yr}^{-1}$}]}

\newcommand{\msim}{\raisebox{-.4ex}{$\stackrel{>}{\scriptstyle \sim}$}}

\newcommand{{\targetname}}{HD 54879}
\newcommand{\new}{}

\authorrunning{Oskinova, Bulik, Nebot}
\titlerunning{Infrared outbursts as potential tracers of common envelope events 
in high-mass X-ray binary formation}
\usepackage{color}

\begin{document}

   \title{Infrared outbursts as potential tracers of common-envelope events 
in high-mass X-ray binary formation}

\author{Lidia M.\ Oskinova\inst{1,4}, 
            Tomasz Bulik  \inst{2,4}, Ada Nebot G\'omez-Mor\'an\inst{3}    }
          
   \institute{\inst{1}{Institute for physics and astronomy, University of 
Potsdam, 
              Karl-Liebknecht-Str. 24/25, D-14476 Potsdam, Germany}\\
              \email{lida@astro.physik.uni-potsdam.de}  \\    
              \inst{2}{Astronomical Observatory, University of Warsaw, Aleje 
Ujazdowskie 4, 00478
Warsaw, Poland} 
\\
              \inst{3}{Universit\'e de Strasbourg, CNRS, Observatoire 
astronomique 
de Strasbourg, UMR
7550, F-67000 Strasbourg, France}   
  \\
\inst{4}{Kavli Institute for Theoretical Physics, University of California, 
Santa Barbara, CA 93106, USA}
   \date{Received ? / Accepted ?}
}

\abstract
{Classic massive binary evolutionary scenarios   predict that a 
transitional 
common-envelope (CE) phase  could be preceded as well as succeeded by the 
evolutionary stage when a binary consists of a compact object and a massive 
star, that is, a\ high-mass X-ray binary (HMXB). The observational manifestations of 
common envelope are poorly constrained.  We speculate that its ejection might 
be observed in some cases as a transient event at mid-infrared 
(IR) wavelengths.} 
{We estimate the expected numbers of CE ejection events and HMXBs per star 
formation unit rate, and compare these theoretical estimates with 
observations.} 
{We compiled a list of 85 mid-IR transients of uncertain 
nature detected by the Spitzer Infrared Intensive Transients Survey and searched 
for their associations 
with X-ray, optical, and UV sources.} 
{Confirming our theoretical estimates, we find that only one potential HMXB  
may plausibly associated with an IR-transient and tentatively propose 
that X-ray source NGC\,4490-X40 could be a precursor to the SPIRITS\,16az event.
Among other interesting sources, we suggest that the supernova
remnant candidate [BWL2012] 063 
might be associated with SPIRITS 16ajc. We also find that two SPIRITS events are 
likely associated with novae, and seven have potential optical counterparts.}
{The massive binary evolutionary scenarios that involve CE events do not contradict currently available observations of IR transients and HMXBs in 
star-forming galaxies.}


\maketitle

\section{Introduction}
\label{sec:introduction}

Massive stars are born and evolve in binaries, often as part of a 
higher-order hierarchical system \citep[e.g.,][]{Pac1971,Van1998}. The binary 
components can expel and exchange mass, can merge, or become unbound. 
Stellar and binary evolution are strongly affected by these processes. 
Among a large variety of possible evolutionary scenarios, we can distinguish three key stages. In the first stage, both components are 
non-degenerate stars, while in the next evolutionary stage, one of the binary 
components collapses to become a neutron star (NS) or a black hole (BH). The 
accretion of matter lost by the remaining non-degenerate star onto an
NS or BH can power strong X-ray emission. Such systems are usually 
observed as high-mass X-ray binaries (HMXBs). In the final evolutionary stage, 
both binary components are degenerate and form a relativistic binary 
\citep[][and references therein]{Postnov2014}. 

While the physics of binaries in each evolutionary stage is intensively studied 
and is reasonably well established, the transitions between evolutionary stages 
are still poorly understood.  Among the major uncertainties in massive binary 
evolution is a transitional phase during which binary components are embedded 
in a shared or common envelope (CE) \citep{Pac1976}. Such a phase occurs when 
a much more massive star fills its Roche lobe and starts to transfer mass onto
the less massive companion. The orbit shrinks and  the mass transfer is unstable.
 During this phase, the 
friction between the stars and the surrounding envelope moves the stars closer 
to each other. The resulting surplus of the orbital energy is eventually spent to 
disperse the CE and eject its material into the interstellar medium (ISM) 
\citep{Iben1993}. 

Although quantitative arguments on the CE physics are broadly discussed 
\citep[see review by][]{Ivanova2013}, there is a dearth of quantitative studies 
capable to make exact predictions on the observable properties of CEs and their 
final ejections \citep[][and references therein]{MacLeod2017, Iaconi2017}. Recent studies have suggested that the CE phase in some binaries could end with the CE ejection, which might be observable as an IR transient \citep{Blag2017}.  

The ``standard scenario'' of massive binary evolution from a main-sequence 
star to a relativistic binary \citep{Heuvel1973, Tut1973} predicts that the CE 
also occurs in the stage when the primary star has already become a compact 
object.  The evolutionary chain of events runs through the stage of an HMXB 
powered by the matter accretion from the secondary star onto the compact object 
\citep[see, e.g., Fig.\,7 in ][]{Postnov2014}. When the secondary fills its 
Roche lobe, it engulfs the compact object, forming a CE. In this phase, the 
accretion of the CE material onto a BH is unavoidable, and the system should be 
visible as a strongly obscured X-ray source. The CE is eventually
expelled,
and the orbit shrinks by a factor of up to a hundred. 
If a merger of the compact object and the secondary star's core is 
avoided, a binary is formed that consists of a He-star and a compact object on a close 
orbit. Such a binary is an HMXB that is powered by wind accretion. This scenario 
is a required ingredient in the formation of merging BHs   similar to GW\,150914 \citep[see Fig.\,1 
in][]{Bel2016}.

There may be observational support for this scenario. Recently, a new 
class of obscured HMXBs was identified.  In these systems, a large amount of 
circumstellar material effectively absorbs X-rays from an accreting compact 
object \citep{Hynes2002, Filliatre2004, Servillat2014, Lau2016, Lau2017}. The 
donors in these objects 
are likely sgB[e] stars \citep{Lamers1998}. We may speculate  that sgB[e] 
HMXBs roughly correspond to the systems where a compact object orbits
a massive star core that is surrounded by a CE or its remnant material. 


In this Letter we take a simple Ansatz that some  IR transients are related 
to the ejection of the CE. We estimate the expected numbers of CE 
events and HMXBs in a star-forming galaxy and compare these estimates with 
observations by searching for possible X-ray counterparts to the IR transients. 

The time is now well suited to investigate the associations between 
IR-transients and X-ray sources. Powerful X-ray observatories, \cxo\ and \xmm, are operating 
since 2000. Although neither  of these observatories has conducted an 
all-sky survey, their respective source catalogs are already very sizable.
Since 2014, the {\em Spitzer} IR telescope conducts a systematic search 
for mid-IR transients in nearby galaxies, the  SPitzer InfraRed 
Intensive Transients Survey (SPIRITS) \citep[see the survey overview in][]{Kas2017}.   
The survey has identified  a new type of events whose nature is not yet known. 
These events  have infrared luminosities between novae and supernovae and occur 
in star-forming galaxies. Their unknown nature and association with 
star-forming regions opens a room for a suggestion that some of these events 
may be associated with a CE ejection. In this Letter we test this hypothesis.

In Section\,\ref{sec:rate} we use basic binary evolution considerations to 
predict 
the number of CE events and HMXBs in a star-forming 
galaxy. Section\,\ref{sec:sample} describes the work done to search for 
correlations between IR-transients and X-ray sources. The conclusions are 
presented in sect.\,\ref{sec:con}. The notes on individual objects and the 
table summarizing our catalog searches are presented in the appendix. 

\section{Expected number of CE events}
\label{sec:rate}

In order to estimate the number of CE events in a star-forming galaxy, we must 
consider the binary evolution leading to such phases. 
{\new We choose a qualitative approach based on making physically plausible assumptions 
rather than a full population synthesis modeling. The key goal of this work is to check 
whether our basic understanding of massive binary evolution contradicts currently 
available observations of star-forming galaxies. }

The  initial state of the binary is determined by the mass of the primary 
$M_{1i}$, the mass ratio $q=M_{2i}/M_{1i}$, and the orbital separation 
$A_i$. We assume for simplicity that the initial orbits are circular.

We assume that the distribution of the primary mass is given by the 
initial mass function (IMF) $\zeta(M)$ \citep{Kroupa1993} with the lowest 
stellar mass $M_{\rm min} =0.08 M_\sun$, and the maximum stellar mass $M_{\rm 
max} =100 M_\sun$.
{\new  This assumption on the upper stellar mass is conservative. 
In  very massive star clusters, the most 
massive stars could have masses of at least  $150\,M_\odot$ 
\citep{Figer2005,Weidner2006}. Moreover, the IMF may depend on metallicity 
becoming more  top-heavy with decreasing metallicity \citep{Marks2012}; this
suggestion is gaining observational support \citep{Schneider2018, 
Ram2018}.}

Furthermore, we assume that the distribution of mass ratios is $\Psi(q) \propto 
{\rm const}\,k$ \citep{Kobul2007},  and the distribution of the initial 
separations is $\Xi(A) \propto A^{-1}$, and we assume that the range of 
the orbital separations is $10R_\sun < A < 10^6 R_\sun$. The range of the mass 
ratio takes into account
the fact that the minimum mass of a star is $M_{\rm min}=0.08 M_\sun$, 
so that for a given mass of the primary $M,$ we have 
$q_{\rm min}(M) = 0.08 M_\sun/M$. We consider two cases for the evolution: 
the high and low initial mass ratio.  

We first consider the case when  the initial mass ratio is low, for instance, $q = q_{\rm div}$, where $q_{\rm div} \lessapprox 0.25$ 
\citep[for a discussion, see, e.g.,][]{Belczynski2008}. 
As the primary evolves, it will fill the Roche lobe as long as 
the initial orbital separation is not too large, $A~< 1000 R_\sun$.
Since the mass ratio is low, the mass transfer will be unstable and can be 
considered a CE event. As a result of the CE, the binary can merge  if the 
initial orbital separation is small, $A~< 100R_\sun$, otherwise it will survive. 
If it survives, the primary will be stripped of its envelope and may become  
a helium star, while the secondary will be essentially unchanged 
because it cannot accrete much matter in the short timescale of the CE 
event. The orbital separation will be greatly reduced \citep{Webbink1984}. In 
this case, we do not expect an HMXB, neither preceding  nor following the CE event.

In the case of a high mass ratio, where the masses are nearly equal, $q\gtrapprox q_{\rm 
div}$, the evolutionary scenario is different. The system enters into the first 
mass transfer that can be considered as non-conservative. It is initially 
unstable, and then stabilizes as the mass ratio is reversed. The orbit 
initially shrinks, but then expands, and the final orbital separation 
is close to the initial separation, $A_{\rm ps}\approx A_i$ \citep{Eggleton2006book}. 
The system now consists of a core of the primary, possibly with some hydrogen 
envelope, and a rejuvenated secondary with an increased mass.  The next stage 
of the evolution is due to the fast evolution of the core of the primary, which 
quickly explodes as a supernova, forming an NS or a BH. Formation of an NS is 
usually associated with a kick \citep{Cordes1998,Hobbs2005}. As the system is 
still relatively wide, the kick can easily disrupt the system, and we assume that 
the survival probability of the system is $p_{\rm SN1}\approx 0.1$. 

{\new Recent modeling has shown that the high Galactic latitudes of many 
black holes in low-mass X-ray binary systems could be explained only if 
these systems did obtain quite significant natal kicks, similar to those of 
NSs \citep{Repetto2012}. On the other hand, the gravitational wave observations 
constrain the kicks  that BHs receive  in high-mass mass binaries to about  
$\sim 50 - 200 $\,km\,s$^{-1}$, that is,\,  smaller than the NS kicks 
\citep{Wys2018}. The relatively small kicks imply a high survival 
probability of massive binaries after the primary collapse into a BH. Here, we 
assume for simplicity that  
the survival probability for BH systems is close to unity, $p_{\rm SN1}\approx 
1$ 
\citep[see also ][]{Belczynski2008}. In any case, as can be seen from Eq.\,(\ref{eq:nhmxb}), the predicted number of HMXBs is directly 
proportional to  $p_{\rm SN1}$ and can therefore be easily corrected for lower 
values.  }

The supernova explosion affects the orbit size and introduces some ellipticity 
because of the mass loss and the kick. The system now consists of a compact object 
and a 
massive secondary, and it likely becomes an HMXB, as the 
massive secondary will
have a strong stellar wind that will lead to accretion onto the compact object.
The secondary star gradually increases its radius, fills the Roche lobe, and 
initializes rapid mass 
transfer 
onto the compact object. The mass transfer will now be unstable
and can be described as a common envelope. If the orbital separation is small, 
the compact object will plunge into the donor and form a Thorne-Zytkow object 
\citep{ThorneZytkow1975} or explode as a $\gamma$-ray burst \citep{Fryer1998}.
{\new We make a crude assumption that the system will survive the CE if the 
orbital separation is $A_{\rm ps} \msim 100\,R_\sun $  (e.g.,\ see 
Fig.\,13 in \citet{Terman1995} and the discussion in Section 3.6 in 
\citet{Postnov2014}.)}

We optimistically assume that the system survives the CE with the donor on the 
Hertzsprung gap \citep{Dominik2012}. 
Following the CE, the system will consist of a compact object in a tight binary 
with the core 
of the donor; most likely a helium  star. The compact object will accrete
matter either through the wind or through Roche-lobe overflow. Thus in this 
case, the CE event will also be followed by the formation of a new HMXB. 

We calculate the rate of CE events in both scenarios. 
We assume that the star formation rate (SFR) is constant and 
denote the binary fraction as $f_{\rm bin}$.
The average mass of  a single star is $\left< M\right>_{\rm sin}=\int M \zeta(M) 
{\rm d}M $,
and the average mass of a binary is $\left< M\right>_{\rm bin}=\int \int M(1+q) 
\zeta(M)\Psi(q) {\rm d}M {\rm d}q $.
The average mass of a star in a population is then
\begin{equation}
M_{\rm ave}=
(1+f_{\rm bin})^{-1}
((1-f_{\rm bin})\left< M\right>_{\rm sin}+
f_{\rm bin}\left< M\right>_{\rm bin}).
\end{equation}
Assuming $f_{\rm bin}=0.5$, and $k=0,$ we obtain $M_{\rm ave}=(5/6)\left< 
M\right>_{\rm sin} =0.37 M_\sun $.
The binary formation rate is  $R_{\rm bin} =  {\rm SFR\,} \times f_{\rm bin} 
M_{\rm ave}^{-1}=1.35({\rm SFR}/M_\sun {\rm yr}^{-1} ) {\rm yr}^{-1}$.

In the  first case, when $q<q_{\rm div}$ , the  rate of the CE events can be 
obtained by 
integrating the relevant distributions:
\begin{equation}
{\cal N}_{\rm CE1} = R_{\rm bin}   f_A
\int_{M_{\rm min1}}^{M_{\rm max}} \zeta(M)  {\rm d}M 
\int_{q_{\rm min}(M)}^{q_{\rm div}} \Psi(q) {\rm d}q, 
\label{eq:NCE1}
\end{equation}
where $M_{\rm min1}$ is the lowest mass of the primary that we are interested 
in,
$f_{A}=\int_{A_{\rm min}}^{A_{\rm max}} dA \Xi(A)$ is the fraction of systems 
with an initial orbital separation 
in the range between 
$ A_{\rm min}=100 R_\sun$, and $A_{\rm max} =1000R_\sun$. Given the above 
assumptions of
the range for $A,$  we obtain $f_A=0.2$.
The initial separations smaller than $A_{\rm min}$ will lead to a merger,
while in the case of initial separations larger than $A_{\rm max}$ , the primary 
will not 
fill the Roche lobe.
The value of $M_{\rm min1}$ can be obtained from the age of the population 
that is analyzed, as we consider only the stars that could have evolved within a 
given
timescale.   For a typical Milky Way-like galaxy, we use $M_{\rm min1} = 1 
M_\sun$.

In the second case, when $q>q_{\rm div}$ , the CE events rate is given by 
\begin{equation}
{\cal N}_{\rm CE2} = R_{\rm bin} p_{\rm SN1} f_A
  \int_{M_{\rm CO}}^{M_{\rm max}} \zeta(M) {\rm d}M 
  \int_{q_{\rm div}(M)}^{1} \Psi(q) {\rm d}q ,
\label{eq:NCE2}
  \end{equation}
where ${M_{\rm CO}}=8 M_\sun$ is the minimum {\new initial}  mass required to 
form a compact object: an NS, or a BH \citep{Woosley2002}.  
Incidentally, we can obtain the expected number of HMXBs
that precede the CE phase in this case. To this end, we  estimate the 
time that each binary is in the X-ray phase as  roughly half of the lifetime
of the secondary after rejuvenation. {\new Given the short massive star 
lifetime, this crude assumption does not introduce large errors.}

We assume that the mass of the 
secondary increases by half in the mass transfer  and becomes 
$M_{\rm 2p}=1.5 M_{\rm 1i} q$. 
The X-ray phase duration is then 
$ T_{\rm X}(M,q) = 0.5T_\star(M_{\rm 2p})
\approx 5.7\times 10^6 (10 M_\sun/(qM_{\rm 1i}))^{2.5}\, \rm{years} $, and the 
number of HMXBs is
\begin{equation}
{\cal N}_{\rm HMXB} =  R_{\rm bin} p_{\rm SN1} f_A
 \int_{M_{\rm min}}^{M_{\rm max}} \zeta(M) {\rm d}M
  \int_{q_{\rm div}(M)}^{1} \Psi(q)T_{X}(M,q) {\rm d}q .
\end{equation}
We can also calculate the number of HMXBs per unit  of star formation:\begin{equation}
 {{N}_{\rm HMXB} \over {\rm SFR}  } = 
 {f_{\rm bin}   p_{\rm SN1} f_A \over M_{\rm ave}}         
  \int_{M_{\rm min}}^{M_{\rm max}} \zeta(M) {\rm d}M 
  \int_{q_{\rm div}(M)}^{1} \Psi(q)  T_{X}(M,q) {\rm d}q 
.\end{equation}
Inserting the numbers, we obtain
\begin{equation}
{{N}_{\rm HMXB} \over {\rm SFR}  }=384 {f_{\rm bin}\over 0.5}{f_{\rm A}\over 
0.2}{p_{\rm SN1}\over 0.1} {\rm yr}M_\sun^{-1}.
\label{eq:nhmxb} 
\end{equation}

To roughly estimate the X-ray luminosity of HMXBs, we assumed that these systems 
are wind fed.  {\new We ran a set of 30\,000 simulations on a grid of plausible  
stellar wind and binary parameters. Circular orbits with orbital separations 
 in the range 2--20\,$R_\ast$ were assumed. Our model population consists of 
binaries with donor masses in the range 15--45\,$M_\odot$ and compact object 
masses in the range 1.4--14\,$M_\odot$ (the donor and compact object masses are 
not drawn from a mass function, but are located on a grid). For each donor star , we 
varied stellar wind parameters. Mass-loss rates in the range 
$\log{\dot{M}}=-7.5\,...\,-5.5\,[M_\odot\,{\rm yr}^{-1}]$ are assigned to a 
donor of each mass. The rationale is that  helium  stars of relatively low present-day mass (such as Wolf-Rayet stars) could have prodigious 
mass-loss rates, while even quite massive stars could have  low mass-loss rates 
at  metallicities lower than solar 
\citep{Bouret2003,Hainich2014,Shenar2016, Hainich2018}. Terminal wind 
velocities in the range 500\,km\,s$^{-1}$ to  2500\,km\,s$^{-1}$ were adopted;  
at close binary separations ($< 5\,R_\ast$), the wind velocity was roughly 
assumed to be a half of its terminal 
 value  \citep{Sander2018}. X-ray luminosities were calculated using   
the Bondi-Hoyle-Lyttleton approximation and accounting for relative 
velocities \citep[using the same formalism as in ][]{Osk2012}.

\begin{figure}[t]
\centering
\includegraphics[width=0.9\columnwidth]{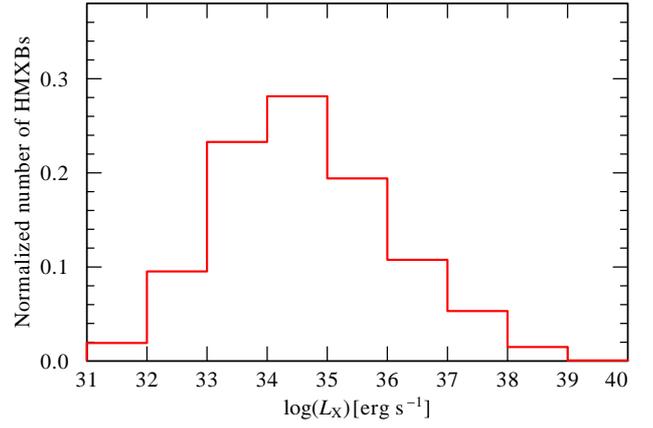}
\caption{Model X-ray luminosity  probability distribution of  
wind-accreting HMXBs (see text for the model description).} 
\label{fig:xlf}
\end{figure}

The histogram  shown in Fig.\,\ref{fig:xlf}  demonstrates that the X-ray luminosity 
of a wind-fed HMXB is likely to exceed $10^{34}$--$10^{35}$\,erg\,s$^{-1}$, 
which is in good agreement with observations \citep{Lutovinov2013}. 
However, our model formalism neglects effects related to the neutron star spin 
and magnetic filed \citep[e.g., ][]{Illarionov1975,Bozzo2016} as well as the 
orbit eccentricity. These effects may lead to temporal arrest of accretion. } 
It  is known that the majority of observed HMXBs are transients 
\citep[e.g.][]{Walter2015,Silvia2017}. Taking 
this into account, the number of HMXBs estimated in Eq.\,(\ref{eq:nhmxb}) is 
in quite good agreement with empirically derived from the observations of 
X-ray sources in nearby galaxies \citep{Mineo2011,Mineo2012}. 

The SPIRITS survey has monitored  191 galaxies \citep{Kas2017}. We do not have 
the estimate of the SFR for each of them, therefore we assume that 
they are similar to the Milky Way. {\new The average SFR of the 
Milky Way varies in the literature from $5\,M_\sun {\rm yr}^{-1}$ to 
$0.7\,M_\sun {\rm 
yr}^{-1}$ \citep{Smith1978,Misiriotis2006,Diehl2006,Murray2010,Rob2010}.
Adopting an SFR $3.6 M_\sun {\rm yr}^{-1}$ per galaxy, }
the SPIRITS survey has examined a region with a total star 
formation rate 
of ${\rm SFR} = 687 M_\sun {\rm yr}^{-1}$. We now evaluate the expected 
rate of CE events 
in the low and high mass ratio case. 
Inserting the fiducial numbers in Eq.\,(\ref{eq:NCE1}), we obtain
\begin{equation}
{\cal N}_{\rm CE1} = 3.0 {f_{\rm bin}\over 0.5}{{\rm SFR}\over 678M_\sun {\rm 
yr}^{-1}} 
{f_{\rm A}\over 0.2} {\rm yr^{-1}}
,\end{equation}
for which we do not expect an X-ray source to precede or follow the CE.  
For the channel with a high mass ratio, where the CE events are associated 
with an X-ray source, we obtain from Eq.\,(\ref{eq:NCE2})
\begin{equation}
{\cal N}_{\rm CE2} = 0.03 {f_{\rm b}\over 0.5}{{\rm SFR}\over 678M_\sun {\rm 
yr}^{-1}} 
{f_{\rm A}\over 0.2}{p_{\rm SN1}\over 0.1}   {\rm yr^{-1}}.
\end{equation}
Thus we expect  that the SPIRITS survey sees only few events from the 
evolutionary channel
with a low initial mass ratio, and there 
is a  very small chance {\new to observe} a CE event associated with a HMXB. 

\section{Search for associations between mid-IR transients and X-ray sources}
\label{sec:sample}

To compare these estimates with observations, we compiled a list of SPIRITS 
events reported in the
Astronomer's Telegrams (6644, 7929, 8688, 8940, 9434, 10171, 10172, and 10488) 
and   in \citet{Kas2017}. The events of known nature 
are excluded from consideration, the remaining 85 IR events are 
listed in Table\,\ref{tab:spirits}. To search for X-ray sources 
that might be associated with SPIRITS events, as a first step, we conducted a blind 
search 
of X-ray catalogs allowing a generous cross-correlation radius. This search 
showed that the absolute majority of X-ray observations of galaxies hosting 
SPIRITS were made by the \cxo\ X-ray telescope. Hence, we 
decided to concentrate on the \cxo\ data as they provide positional accuracy 
compatible with that of the {\em Spitzer} IR telescope. In addition
to searches in the 
catalogs of individual galaxies (when available), we also included the most 
recent meta-catalogs such as the ``X-ray emission from star-forming galaxies - I. 
High-mass X-ray binaries'' \citep{Mineo2012}, 
``The \cxo\ ACIS Survey of X-ray Point Sources in 383 Nearby Galaxies'' \citep{Liu2011},
and ``The \cxo\ ACIS Survey of X-Ray Point Sources: The Source Catalog'' \citep{Wang2016}.

We investigated the coverage of SPIRITS sources by X-ray observations. Of 85 SPIRITS IR-transients, only 7 were not in the 
field of view of \cxo\ or \xmm\ observations. Using the \cxo\ science archive 
facility, we roughly estimated the total \cxo\ exposure time for each SPIRITS 
event 
(Table\,\ref{tab:spirits}).  The 
exposure times are vastly different, and the upper limits for the potential 
X-ray counterparts of X-ray transients are not uniform. It is beyond the scope 
of this study to derive the upper limits on the X-ray non-detections, as we 
are mainly interested in finding positive matches.  

As a next step, the cross-correlations between IR and X-ray sources were 
searched within a radius of $1''$ . 
If an X-ray counterpart was suspected, the X-ray images and event lists were 
retrieved and checked manually. This narrowed search returned
an X-ray counterpart 
for only one object,  SPIRITS\,14ajc in the galaxy M\,83 \citep{Kas2017}. 
Relaxing the cross-correlation radius  to $2''$ added X-ray sources that might be 
associated with SPIRITS\,17mj in M\,81 and SPIRITS\,16az in the galaxy 
NGC\,4490. These and other interesting objects are discussed in the appendix.

{SPIRITS\,16az} is the best potential match we have found.  It was 
discovered on 
\textcolor[rgb]{0,0,0}{ 
\textcolor[rgb]{0.988235,0.501961,0.0313726}{\textcolor[rgb]{0,0,0}{2016-3-5}} 
}in the galaxy NGC\,4490. SPIRITS\,16az has a likely 
optical counterpart, XMMOM J123027.7+413943, at $0\farcs7$ distance with a positional error of 
$0\farcs6.$  The optical source was detected in 2004.
No X-ray source is seen in early $\sim 20$\,ks \cxo\ observations taken on 
2000-11-03. In 2004, two \cxo\ observations were obtained about four months 
apart. On 2004-07-29, no X-ray source in the vicinity of 16az is seen (see left 
panel in Fig.\,\ref{fig:16az}). On the other hand, on 2004-11-20, an X-ray 
source 
is detected about $1''$ away from 16az. This is significant, since both \cxo\ 
observations had similar exposure times of $\sim 40$\,ks. The X-ray source seen 
on 2004-11-20, NGC\,4490-X40, is listed in the \cxo\ catalogs by 
\citet{Mineo2012, Liu2011} and \citet{Wang2016}. With a $1''$ positional error 
on the X-ray source, the association between NGC\,4490-X40 and 16az is 
plausible. The X-ray luminosity of NGC\,4490-X40, $L_{\rm X}\approx 4\times 
10^{37}$\,erg\,s$^{-1}$ at 7.8\,Mpc, is compatible with its being an HMXB.  
Further studies of NGC\,4490-X40 are needed to unambiguously conclude whether 
this source is an HMXB and whether {SPIRITS\,16az} was a CE ejection event.

In Table\,\ref{tab:spirits} we compile the results of our study. Although we 
were not 
successful and did not find an unambiguous association between X-ray sources 
and IR-transients, the estimates presented in section\,\ref{sec:rate} show that 
the one possible X-ray counterpart to an IR-transient is in agreement with the
expected rate of CE ejections linked to the HMXB evolution for a sample of 
surveyed galaxies.

\section{Summary and conclusions}
\label{sec:con}

Motivated by massive binary evolutionary scenarios that predict links between   
HMXBs and short CE events, we searched for X-ray counterparts of IR-transients. 
This was done by correlating the positions of 85 not yet 
identified transients observed by the {\em Spitzer} IR telescope 
with X-ray catalogs and images.  We also checked available optical and UV 
catalogs to confirm an HMXB nature of any potential X-ray counterpart.
We found potential pre-IR outburst optical counterparts, including H\,{\sc ii} 
regions, for seven SPIRITS events. While confirmations for these 
identifications are required, at least some IR-transients we 
consider here are probably linked to young massive stars. 
Two IR-transients in our sample of 85 objects are very plausibly associated 
with recent novae. 
We did not find an unambiguous HMXB counterpart to any of 
the IR-transients. Our best match is SPIRITS 16az, which might
be linked 
with an optical source and an X-ray transient identified as the HMXB 
NGC\,4490-X40. 
The relatively large positional error of the X-ray source prevents firm 
identification.  
Another interesting source is SPIRITS 16ajc, which is likely associated with a 
pre-outburst X-ray source, possibly the SNR candidate [BWL2012] 063. 

From considerations of binary evolution, we 
estimated the expected numbers of CE events and HMXBs in star-forming 
galaxies. We conclude that

\smallskip
\noindent
1) massive binary evolutionary scenarios predict that a CE stage 
can be immediately preceded and/or  succeeded by an  HMXB stage. 
Assuming that a CE ejection could be observed as an IR outburst, in principle 
some IR-transients could  have an X-ray counterpart prior to or post IR-outburst; 

\smallskip
\noindent
2) our estimates show that a few hundred HMXBs per one star formation rate unit 
could be expected in a galaxy. The good agreement between observed numbers 
of HMXBs in star-forming galaxies and these predictions provides a 
justification for our estimates of the expected CE events. Assuming that the 
visibility of CE ejection event is about one year, we estimate that there 
is $\sim 3$\%\ chance to observe an associated IR transient in a sample of 
191 galaxies observed in the SPIRITS survey, if each of these galaxies has an SFR 
similar 
to that of the Milky Way. Hence the expected rate of IR transients associated 
with CE ejection is $\approx 1.6\times 10^{-4}$\,yr$^{-1}$ per Milky Way-type 
galaxy;

\smallskip
\noindent
3) the current lack of positive detections of a CE event 
associated with a massive X-ray binary does not contradict
standard massive binary evolutionary scenarios.

\begin{acknowledgements}
The authors thank the referee for their very useful comments. 
This work has been financially supported by the Programme National Hautes
Energies (PNHE), project 994584. LO and AN thank {\em Integrated Activities in 
the High Energy Astrophysics Domain} project for support that enabled this 
work. 
L.O. acknowledges support from the DLR grant 
50 OR 1612. TB was supported by the FNP grant TEAM/2016-3/19. LO and TB thank 
Kavli Institute for Theoretical Physics, University of California, where this 
project have been initiated, for their support. This research has made use of 
the SIMBAD database,  the VizieR catalogue access tool, operated at CDS, 
Strasbourg, France, and the High Energy Astrophysics Science Archive Research 
Center (HEASARC), which is a service of the Astrophysics Science Division at 
NASA/GSFC and the High Energy Astrophysics Division of the Smithsonian 
Astrophysical Observatory. The scientific results reported in this article are 
based to a significant degree on the data obtained from the Chandra and {\em 
XMM-Newton} Data Archives.
\end{acknowledgements}
%
%
%

\begin{appendix} 

\section{Notes on some interesting individual objects}
\label{sec:io}

\begin{figure*}[th!]
\centering
\includegraphics[width=1.75\columnwidth]{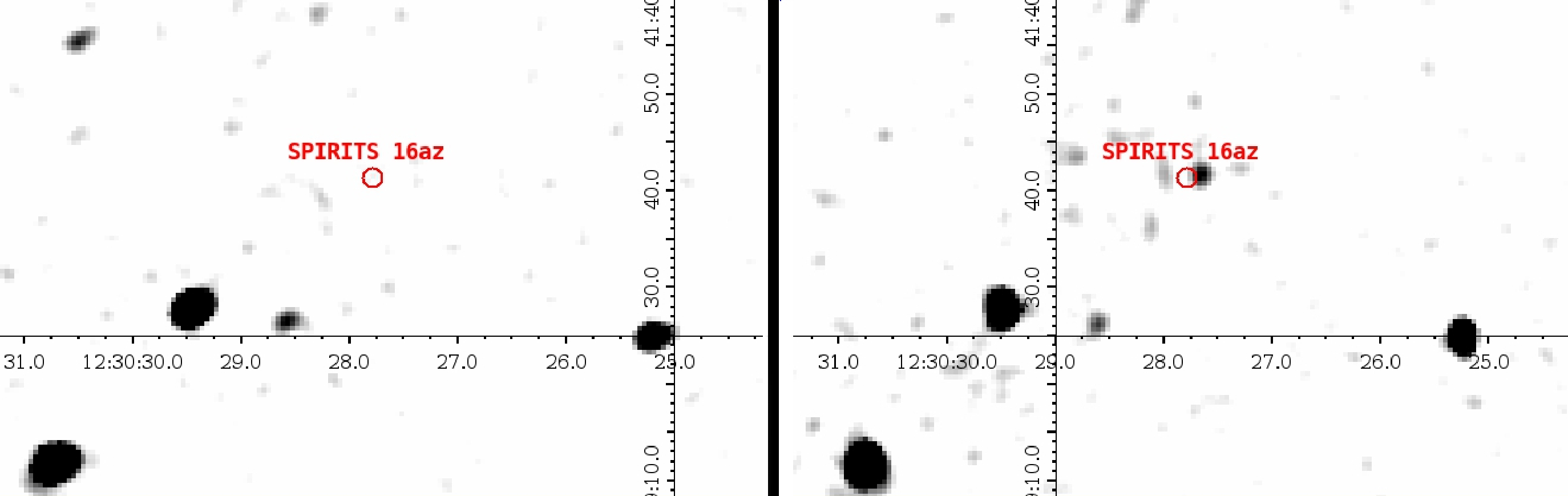}
\caption{Archival not-binned broad-band (0.2-12\,keV) X-ray images of 
NGC\,4490 around the position of SPIRITS\,16az as indicated by the 
red cicles in each panel. The circle radii are $1''$. The 
images were obtained by the \cxo\ ACIS-S camera on  2004-07-29 with an 
exposure time $\sim 39$\,ks (left panel) and  on 2000-11-03 with an 
exposure time $\sim 40$\,ks (right panel). The X-ray images are shown on linear 
scale
and were smoothed to facilitate perception of the source. North is up, east is to 
the left.} 
\label{fig:16az}
\end{figure*}

{SPIRITS\,16ajc} in the galaxy M\,83 is extensively discussed by 
\citet{Kas2017}. The source 
went into IR-outburst in 2010. No optical or near-IR counterpart was detected 
in ground-based follow-up in 2014, but shock-excited H$_2$ emission lines 
were seen in the spectrum measured on the 2014 June 8 spectrum. Our search for 
X-ray sources associated with IR-transients revealed that 
the X-ray source [LKB2014] X100 in the ``M 83 Chandra X-Ray Point Source 
Catalog'' \citep{Long2014} is  $0\farcs5$ apart from the coordinates of 
SPIRITS\,16ajc. [LKB2014] X100  has an X-ray luminosity of 
$L_{\rm X}=8.5\times 10^{35}$\,erg\,s$^{-1}$ and is identified in the catalog 
as a supernova remnant (SNR) candidate. This identification is based on the 
previous detection 
of an SNR candidate 
[BWL2012] 063 at this location in the optical \citep{Blair2012}. The optical 
observations were 
obtained on 2009 April 26 and 27, i.e., pre-IR outburst. \citet{Kas2017} 
inspected the 
HST WFC3 images of the area around SPIRITS 14ajc taken in 2012 (i.e.,\ during 
the 
IR outburst) 
They noted a faint emission nebula close to 14ajc in H$\alpha$+[N\,{\sc ii}] 
filter image,
but commented that this nebula is outside of the 14ajc position error circle. It is 
possible
that  this faint emission nebula is the SNR candidate, and thus is not 
associated with 14ajc. 

The galaxy NGC\,1313 is a host of a few SPIRITS transients.
{SPIRITS\,16tj} was detected in this galaxy on 2016-08-05 
\citep{Jencson2017}. 
Although the galaxy was observed by both the \cxo\ and \xmm\ telescopes, no X-ray 
source 
potentially associated with 16tj was found  in the X-ray catalogs. 
An optical source likely associated with SPIRITS\,16tj was detected by the \xmm\ 
optical/UV telescope (OM) \citep{Page2012}. The catalog position of the OM UV 
source 
is $0 \farcs 18$ source from 16tj, while its statistical positional uncertainty 
is $0 \farcs 62$. The $B_{\rm Vega}=22.6\pm 0.4$\,mag. The source is associated 
with 
the star cluster [L2004]\,1313-464 with V=19.9\,mag \citep{Larsen2004}. The 
optical 
and UV source is seen off-set by $\sim 0\farcs 2$ from nominal coordinates of  
the 
16tj in the pre-outburst HST images obtained on 2014-02-19. A comparison of 
pre- and post-outburst HST can help to establish whether 16tj was indeed 
associated with an optical star in a cluster.  Optical/UV sources are also seen 
in the vicinity of the {SPIRITS\,16tg} and {SPIRITS\,16tf} events. 
On the other hand, no optical precursor to {SPIRITS\,16th} is obvious in the 
pre-outburst HST images.  

{SPIRITS\,15aht} discovered on 2016-1-14 \citep{Jencson2016} is likely 
associated with the nova candidate PNV J09551857+6904223. The latter has 
H$\alpha$ 
magnitude 
$19.5\pm 0.1$\,mag and was discovered on 2015 Oct. 14.198 UT \citep{Horn2015}. 
This nova  is only $0\farcs5$ away from 15aht. These two events are very 
probably associated. 

{SPIRITS\,15ael} in the dwarf elliptical galaxy NGC\,205  was 
detected only $0\farcs6$ away from the  UV source XMMOM J004022.8+414136  
(positional error $0\farcs53$). \xmm\ observations were carried out on 
2004-01-02, i.e.,\ 
more than a decade prior to the detection of an IR transient. Likely, the IR 
transient and the UV source are associated with the objects cataloged by 
\citet{Lee1996}. Unfortunately, the archival \xmm\ and \cxo\ observations are 
quite shallow.  
 
{SPIRITS\,15ud} in the galaxy M\,100 is probably associated with the 
H\,{\sc ii} region [K98d] 988,  while {SPIRITS\,14bsb} is associated with 
the H\,{\sc ii} region [SCM2003] NGC 625 5 in the galaxy NGC\,625.
 
{SPIRITS\,16az} in the galaxy NGC\,4490 was in the field of view of six \xmm\ 
observations in 2002, 2008, and 2015. In none of these observations
was NGC 4490 
X40 detected as a point source. The reason may be that the object was not 
outbursting during the \xmm\ observations, or because it is located in a 
crowded area filled with diffuse X-rays and thus is difficult to detect with \xmm.

{SPIRITS\,14axa} has peaked at 2014-06-13 \citep{Kas2017}. This transient is 
only $0\farcs44$ away from a nova discovered on 2014 May 21.92 
\citep{Horn2014}. These two events are likely related. 

SPIRITS\,17mj in M\,81 is only $1\farcs4$ away from the X-ray source NGC 
3031-X165 listed in the \citet{Liu2011} catalog of X-ray sources in nearby 
galaxies. However,  no counterparts are found in other X-ray 
catalogs. The visual inspection of \cxo\ X-ray images does not reveal an 
obvious X-ray source at this position (in many observations, the source lies in 
the CCD gap).

\begin{table*}
\scriptsize
\caption{Infrared transients, SPIRITS, considered in this work. \cxo\ exposure 
times shown in col. 6 are approximate and can be 
used only for a rough guidance }          
\label{tab:spirits}
\begin{tabular}{rccccrl}
\hline      
\rule{0pt}{2.2ex} 
\small & RA         &    DEC      & SPIRITS & Host galaxy & CXO exposure time & comments \\
  &         &             &       &           & [ks]   & \\ \hline
1 & 01:35:01.12 & -41:26:07.2 & 17eq  & NGC0625   & 60   &  \\  
2 & 13:25:17.57 & -42:59:21.0 & 17bf  & Cen A     & 1100 &  \\
3 & 03:46:35.25 & +68:01:57.1 & 17c   & IC 342    & 70   &  \\
4 & 03:18:21.22 & -66:29:51.1 & 16tj  & NGC 1313  & 70   & optical/UV src at $0\farcs 2$ \xmm\ OM \& HST, cluster [L2004]\,n1313-464\\ 
5 & 03:18:27.20 & -66:28:38.6 & 16th  & NGC 1313  & 85   & \\
6 & 03:18:16.92 & -66:28:57.7 & 16tg  & NGC 1313  & 85   & UV src at $0\farcs 5$ in the HST image\\ 
7 & 03:18:04.82 & -66:30:17.8 & 16tf  & NGC 1313  & 85   & UV src at $0\farcs 5$ in the HST image\\  
8 & 13:05:43.63 & -49:27:04.2 & 16rz  & NGC 4945  & 450  & \\
9 & 13:05:07.36 & -49:32:40.0 & 16rs  & NGC 4945  & 126  & \\
10 & 13:05:22.33 & -49:28:31.2 & 16rp  & NGC 4945  & 450  & \\
11 & 13:05:19.00 & -49:28:50.8 & 16rn  & NGC 4945  & 450  & \\
12 & 13:36:47.15 & -29:52:58.6 & 16pr  & M 83      & 840  & \\
13 & 13:37:01.50 & -29:54:26.8 & 16po  & M 83      & 840  & \\
14 & 03:46:27.63 & +68:13:42.0 & 16ph  & IC 342    & 70   & \\
15 & 13:36:40.73 & -29:52:41.7 & 16oz  & M 83      & 840  & \\
16 & 13:18:53.54 & -21:04:14.5 & 16oj  & NGC 5068  & 54   & \\
17 & 14:02:57.80 & +54:22:50.6 & 16kp  & M 101     & 800  & \\
18 & 12:50:50.53 & +41:06:04.4 & 16ko  & M 94      & 74   & \\    
19 & 12:40:06.02 & -11:38:14.1 & 17as  &  M 104    & 92   &  $1\farcs 8$ away from the globular cluster [LFB2001] C-101\\
20 & 13:05:36.89 & -49:23:38.2 & 17ar  &  NGC 4945 & 90  & \\
21 & 03:47:29.89 & +68:03:13.1 & 17g   &  IC 342   & 14 & \\  
22 & 00:40:20.95 & +41:39:35.6 & 16abq &  M 110    & 9  & \\ 
23 & 03:46:49.27 & +68:03:52.9 & 17lk  & IC 342    & 72 & \\ 
24 & 03:46:03.55 & +68:08:48.2 & 17lg  & IC 342    & 72 & \\ 
25 & 03:46:07.32 & +68:07:59.0 & 17lc  & IC 342    & 72 & \\ 
26 & 06:16:27.78 & -21:22:51.7 & 17lb  & IC 2163   & 62 & \\  
27 & 13:25:33.08 & -43:00:51.6 & 17kw  & Cen A     & 1100 & \\
28 & 13:25:32.03 & -43:00:41.3 & 17kq  & Cen A     & 1100 & \\
29 & 13:25:20.3  & -42:59:20.2 & 17kp  & Cen A     & 1100 & \\
30 & 13:36:57.42 & -29:50:19.1 & 17kj  & M83       & 810  & \\
31 & 13:25:22.18 & -43:01:17.5 & 17kf  & Cen  A    & 1100 & \\
32 & 13:25:29.44 & -43:01:40.0 & 17kc  & Cen A     & 1100 & \\
33 & 13:37:16.22 & -29:54:18.7 & 17ka  & M83       & 810  & \\
34 & 12:18:50.33 & +47:18:11.5 & 17fo  & NGC4258   & 500  & \\
35 & 14:03:01.29 & +54:22:54.4 & 17fm  & M101      & 885  & \\
36 & 23:57:44.77 & -32:34:58.4 & 17fe  & NGC 7793  & 190  & \\
37 & 09:55:18.54 & +69:04:22.8 & 15aht &  M81      & 800  & $0\farcs 5$ offset from a nova candidate 2015 Oct. 14.198 UT \\
38 & 07:36:37.40 & +65:38:02.6 & 15ahg &  NGC 2403 & 80   & \\
39 & 03:46:17.57 & +68:08:44.7 & 15agl &  IC342    & 30   & \\
40 & 20:34:59.65 & +60:11:18.1 & 15afp &  NGC6946  & 163  & \\
41 & 02:22:40.29 & +42:23:53.5 & 15aev &  NGC 891  & 110  & \\
42 & 00:40:22.82 & +41:41:36.4 & 15ael &  NGC 205  & 9    & optical source $0\farcs6$ V=20.6\,mag\\ 
43 & 15:22:05.55 & +05:03:15.9 & 15ade &  NGC 5921 & not observed & \\
44 & 03:18:23.63 & -66:30:24.2 & 15aag &  NGC1313  & 65  & \\
45 & 13:36:57.08 & -29:53:13.3 & 15aac &  M83      & 810  & \\
46 & 13:34:44.21 & -45:32:25.5 & 15yq  &  ESO270G017 & not observed & \\
47 & 00:54:48.74 & -37:43:13.7 & 15yf  &  NGC300   & 200 & \\
48 & 12:34:19.00 & +06:28:15.7 & 15ue  &  NGC 4532 & not observed & \\
49 & 12:22:55.29 & +15:49:22.0 & 15ud  &  NGC 4321 & 134 & H\,{\sc ii} region \\
50 & 11:18:19.10 & -32:51:06.9 & 15ua  &  NGC 3621 & not observed & \\
51 & 01:35:06.72 & -41:26:13.4 & 14bsb & NGC 625   & 60 & H\,{\sc ii}  region \\
52 & 00:54:49.68 & -37:39:51.2 & 14bmc & NGC300    & 150 & \\
53 & 22:02:41.52 & -51:17:34.5 & 14beq & IC5152    & not observed & no optical counterpart in post-outburst images\\
54 & 12:56:43.25 & +21:42:25.7 & 14bay & NGC 4826  & 27 & \\
55 & 12:29:03.16 & +13:11:30.7 & 16ix  & NGC 4458  & 34 & \\ 
56 & 12:56:39.10 & +21:41:43.2 & 16fz  & NGC 4826  & 27 & \\
57 & 12:15:38.61 & +36:19:46.9 & 16ea  & NGC 4214  & 56 & \\
58 & 12:30:27.78 & +41:39:41.3 & 16az  & NGC 4485/4490 & 78 & optical $0\farcs6$, X-ray at $1\farcs4$, HMXB?\\ 
59 & 09:32:11.64 & +21:30:03.0 & 16aj  & NGC 2903  & 93 & \\
60 & 03:18:09.34 & -66:29:59.4 & 15qv  & NGC1313   & 90 & \\ 
61 & 03:18:15.26 & -66:30:03.4 & 15qo  & NGC1313   & 85 & \\ 
62 & 22:02:42.07 & -51:17:22.3 & 15qh  & IC 5152   & not observed & \\
63 & 10:44:02.36 & +11:42:15.1 & 15pz  & NGC 3351  & 128 &  \\ 
64 & 13:37:08.37 & -29:50:19.7 & 15nz  & M83     & 840 & \\
65 & 12:39:54.87 & +61:36:46.3 & 15mr  & NGC 4605 &   not observed & \\ 
66 & 14:03:10.76 & +54:22:49.1 & 15mo  &  M101 & 850 & \\ 
67 & 14:03:49.44 & +54:20:50.7 & 15mn  &  M101 & 680 & \\
68 & 10:03:18.90 & +68:43:52.1 & 15mk  &  NGC 3077  & 60 & \\ 
69 & 09:55:28.72 & +69:39:58.6 & 14qk  & M82 & 813 & \\ 
70 & 12:50:49.56 & +41:05:52.7 & 14afv & NGC4736 & 71 & \\ 
71 & 13:05:30.87 & -49:26:50.8 & 14agd & NGC 4945 & 449 & \\ 
72 & 13:36:52.95 & -29:52:16.1 & 14ajc & M83 & 840 & X-ray source at $0\farcs5$, SNR? \\
73 & 13:37:05.02 & -29:48:56.2 & 14ajd & M83 & 840 & \\ 
74 & 14:02:55.51 & +54:23:18.5 & 14aje & M101 & 885 & \\
75 & 13:37:12.71 & -29:49:14.9 & 14ajp & M83 & 508 & \\
76 & 13:36:54.81 & -29:52:33.7 & 14ajr & M83 & 840  & \\
77 & 03:47:03.17 & +68:09:05.3 & 14ave & IC342 & 15 \\ 
78 & 09:56:01.52 & +69:03:12.5 & 14axa & M81 & 790 & associated with nova PNV J09560160+6903126\\
79 & 07:36:34.70 & +65:39:22.4 & 14axb & NGC2403 & 38 & \\ 
80 & 13:39:50.99 & -31:38:46.0 & 14bay & NGC5253  & 191 & \\ 
81 & 12:56:47.90 & +21:41:13.4 & 17pe  & M64      & 28  &  \\
82 & 13:18:51.52 & -21:01:35.1 & 17nx  & NGC 5068 & 54  &  \\
83 & 09:55:36.20 & +69:06:21.0 & 17mj  & M81      & 300   & possible X-ray 
source at $1\farcs4$ \\
84 & 07:36:35.94 & +65:37:26.5 & 17mi  & NGC 2403 & 230  & \\
85 & 13:05:11.16 & -49:30:19.0 & 17mb  & NGC 4945 & 300  & \\
 & \end{tabular}                          
        
\end{table*}


\end{appendix}

\end{document}